\documentclass[3p,times,twocolumn,preprint]{elsarticle}

\usepackage{epsfig}
\usepackage{xcolor}
\usepackage{lineno}

\def\um{\ifmmode {\mathrm{\mu m}}\else
                  \textrm{$\mu$m }\fi}%
\def\MeV{\ifmmode {\mathrm{\ Me\kern -0.1em V}}\else
                   \textrm{Me\kern -0.1em V}\fi}%
\def\keV{\ifmmode {\mathrm{\ ke\kern -0.1em V}}\else
                   \textrm{ke\kern -0.1em V}\fi}%
\def\eV{\ifmmode  {\mathrm{\ e\kern -0.1em V}}\else
                   \textrm{e\kern -0.1em V}\fi}%
\def\uW{\ifmmode  {\mathrm{\mu  W}}\else
                   \textrm{$\mu$W}\fi}%

\begin{document}

\begin{frontmatter}

\date{20240131 submitted, 20240428 proof}

\title{
Characteristics of the MTx optical transmitter in
Total Ionizing Dose
}

\author[SMU]{D. Gong}
\author[IPAS]{S. Hou\corref{suen}}
\author[IPAS]{B.J. Juang}
\author[IPAS,NTU]{J.-H. Li}
\author[SMU]{C. Liu}
\author[SMU]{T. Liu}
\author[NJU]{M. Qi} 
\author[SMU]{J. Ye}
\author[NJU]{Lei Zhang} 
\author[SMU]{Li Zhang} 
\author[INER]{H.P. Zhu} 
 
\cortext[suen]{Corresponding author. 
               E-mail address: suen@sinica.edu.tw }

\address[SMU]{
 Southern Methodist University, Dallas, TX 75205, U.S.A. }
\address[IPAS]{
 Institute of Physics, Academia Sinica, Taipei, Taiwan 11529 }
\address[NTU]{
 National Taiwan University, Taipei, Taiwan, 10617 }
\address[NJU]{
 Nanjing University, Nanjing, Jiangsu 210093, China }
\address[INER]{
 Institute of Nuclear Energy Research, Taoyuan, Taiwan 32546 }

\begin{abstract}

The dual-channel multi-mode 850 nm optical Miniature Transmitter (MTx) is developed 
for data transmission of the ATLAS LAr calorimeter readout at LHC.
The MTx’s are exposed to the radiation field of proton-proton collisions, 
therefore, the tolerance in Total Ionizing Dose (TID) is required.
The TID effects in the MTx are investigated with 
X-rays and Co-60 gamma-rays for the active components of 
VCSEL diodes, and the customized {\color{black} Link-on-Chip laser driver (LOCld) } 
developed in 0.25 $\mu$m Silicon-on-Sapphire CMOS technology. 
The irradiation tests were conducted at various dose rates. 
The responses to TID are observed with degradation of laser currents 
at initial dose of 10 to {\color{black} 100~Gy(SiO$_2$)},
and partial recovery with additional TID 
to a stable output about 90 \% of the original.
The optical eye diagrams of irradiated samples show slightly increased jittering, 
and are suitable for the ATLAS requirement of 5~Gbps applications.

\end{abstract}

\end{frontmatter}

\section{Introduction }

\begin{figure}[b!] 
  \centering
    \includegraphics[width=.7\linewidth]{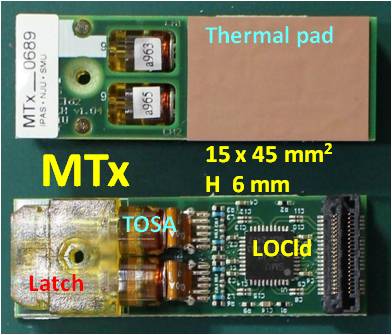}
    \caption{ The MTx module of lateral dimension 15 mm $\times$ 45 mm is
       packaged with a high-density connector 
       (LSHM-120-02.5-L-DV-A-N-TR, SAMTEC) and
       a customized LC fiber connector latch on the TOSAs 
       (TTF-159, Truelight) to a total stack height of 6~mm. 
    \label{fig:MTx} }
\end{figure}

The dual-channel optical Miniature Transmitter 
(MTx) is developed for applications in the
ATLAS experiment at the Large Hadron Collider 
(LHC)~\cite{MTx, MTx_MTRx, MTx_trigger}.
The type of low-height (6 mm) assembly is shown in Fig.~\ref{fig:MTx}.
Near 4k modules are produced for the Liquid Argon 
calorimeter (LAr)~\cite{LAr} and the muon New Small Wheel spectrometer 
(NSW)~\cite{NSW}, for the Phase-I upgrade to enhance event
trigger performance. 
The data transmission is conducted with
the MTx and MTRx modules at 5.12 Gbps.
The MTRx is a single channel MTx assembled with a customized  {\color{black}
``Receiver Optical Sub-Assembly'' (ROSA)}
using a rad-hard GBTIA chip~\cite{GBTIA}.

The MTx has two active devices, which are the 
Vertical-Cavity Surface-Emitting Laser
(VCSEL) diode   packaged in  {\color{black}
``Transmitter Optical Sub-Assembly'' (TOSA), 
and the radiation tolerant 
{\color{black} ``Link-on-Chip laser driver'' (LOCld)}~\cite{LOCld,MTx_prottype}.}
The LOCld is developed with the 0.25 $\mu$m Silicon-on-Sapphire 
(SoS)  CMOS technology.  {\color{black}
The radiation with Total Ionizing Dose (TID) in MOS materials creates
electron-hole pairs, with holes being trapped in 
SiO$_2$ and defects formed at Si/SiO$_2$ interfaces~\cite{Oldham}. 
}

\begin{figure*}[t] 
      \vspace{-3mm}
  \centering
    \includegraphics[width=.72\linewidth]{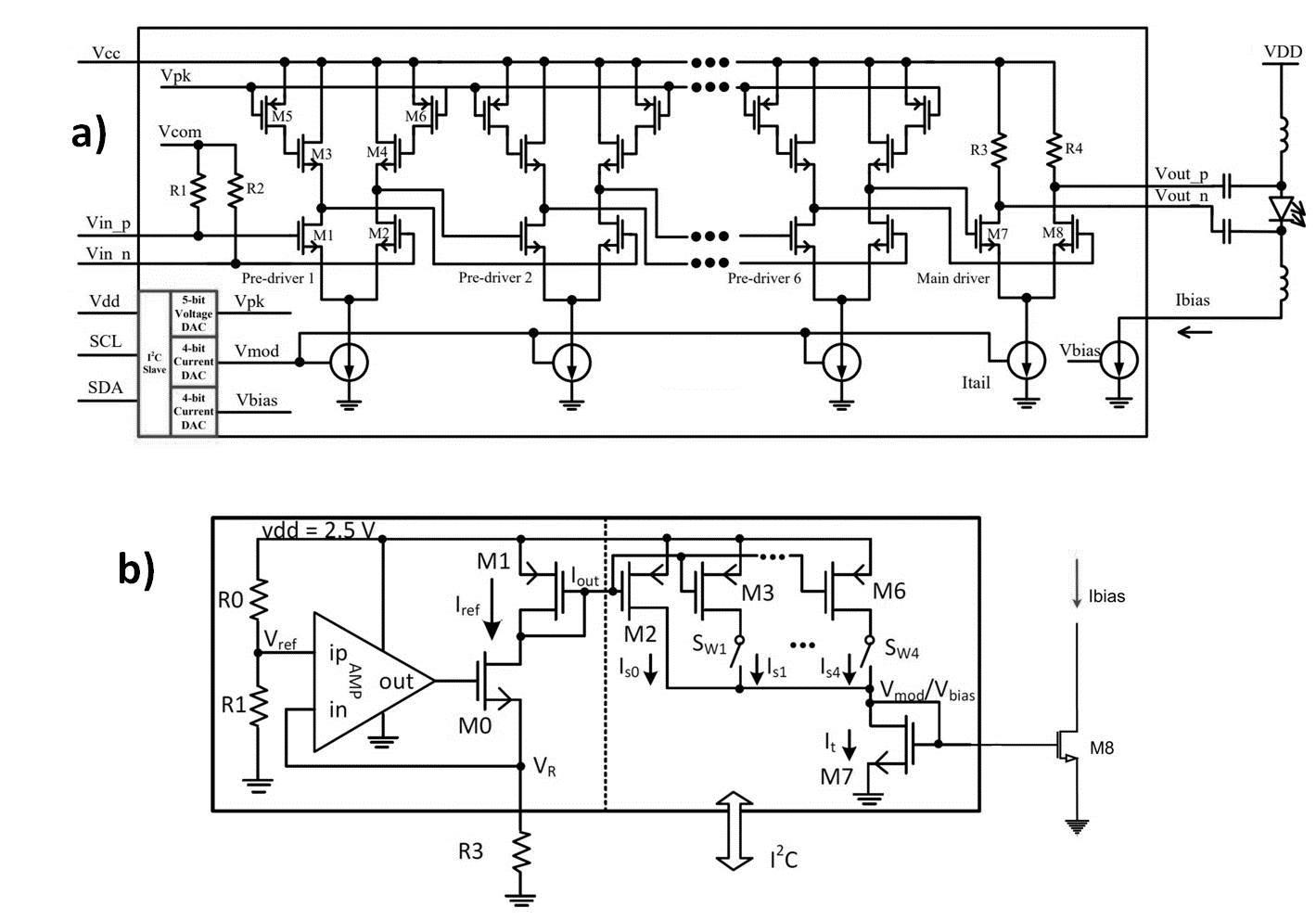}
    \vspace{-4mm}
    \caption{  The circuit design of the dual-channel 
    laser driver LOCld is plotted in a). 
    The Vcc on driver stages and VDD on laser are biased at 3.3~V. 
    The I$^2$C communication and control block shown in the lower left 
    corner is biased with Vdd at 2.5 V. 
    The programmable current to the laser diode is illustrated in b).
  \label{fig:LOCld} }
\end{figure*} 

The MTx modules mounted on the LAr trigger digitizer boards 
are exposed to  the radiation field of $p$-$p$ collisions.
Degradation of the laser light power by 7.5~\%
is observed in the initial operation with $p$-$p$ center-of-mass energy
 $\sqrt{s} = 13.6$~TeV in 2022,
which corresponds to 11 fb$^{-1}$ integrated luminosity\footnote{ \color{black}
  ATLAS LAr Phase I meeting, S. Menke, ``LTDB power loss versus TID'', 
   26, October, 2022.}.
The simulation of radiation field suggests that the TID accumulated is 
{\color{black} 1.5 Gy(SiO$_2$).}
The dose rate is at order of {\color{black} 
0.01~Gy(SiO$_2$)/hr.}

The LOCld prototypes had been evaluated with
X-ray irradiation.
To confirm the TID effects, studies with X-ray and Co-60 irradiations 
were conducted again for the dependence attributed to the VCSEL and 
the LOCld, on the difference of gamma sources, dose rates and 
annealing behaviors.

In Section~2, the MTx assembly and the LOCld laser driver design are 
described.  The X-ray irradiation tests are discussed in Section~3.
The degradation of MTx is examined for the bias currents 
and optical powers versus TID in increasing dose rates to {\color{black}
1.62 kGy(SiO$_2$).}
The effects on  eye diagrams at 5~Gbps are shown.
Section~4 presents the Co-60 gamma-ray tests on VCSEL and MTx samples.
The characteristic changes are compared for the dose rates
and possible annealing behaviors, with the samples irradiated to
various doses of up to {\color{black}
1.15 kGy(SiO$_2$). }

The most profound TID effects on MTx are observed with the degradation 
of LOCld laser currents
that drops by about 15~\% at the initial dose of 10 to {\color{black}
100~Gy(SiO$_2$),}
and recovers with additional TID, to 90~\% of the original. 
The optical eye diagrams show slightly higher 
jitter noise and remain qualified for 8~Gbps data transmission.
A short summary is discussed in Section~5.

\section{MTx assembly with the LOCld ASIC }

The MTx transmitter is designed for applications in the LHC radiation 
environment with more than ten years
in operation~\cite{Opto-radhard, MTx_Aging}.
The module assembly for LAr (Fig.~\ref{fig:MTx}) has
the TOSA packages held by a customized latch, 
and plug-in of fiber-optic cables terminated with LC-type ferrules.
The LOCld is packaged in {\color{black} ``Quad-Flat No-lead'' QFN-40 format}
for assembly on the MTx module 
circuitry suitable for 10~Gbps speed performance.
The schematics of the ASIC is plotted in Fig.~\ref{fig:LOCld}.

The LOCld circuits have two bias voltages, with the 3.3 V 
to power the driver stages and the two laser driver channels,
and the 2.5~V for the I$^2$C communication and control block, respectively. 
The laser driver channels are composed of 6 stages of pre-driver 
and a main driver stage.
Each pre-driver stage imposes active shunt peaking~\cite{peaking} 
to boost the bandwidth, and the peaking strength is programmable. 

All of the pre-driver stages have the same structure 
but different transistor size, with the following pre-driver stage having 
larger transistor size and stronger driving capability. 
The main driver provides 50~Ohm 
termination and programmable output modulation current for the VCSEL diode. 
The biasing current is programmable by a current source connected to 
the cathode of VCSEL diode.
The output current to VCSEL is configured to 6~mA.
The TOSA light output is specified in the range of 0.54 to 1.02~mW.

The prototype LOCld chips were irradiated with X-rays 
in biased condition and revealed degradation of 
laser current by about 7.5~\%~\cite{LOCld_ASIC}.
In Fig.~\ref{fig:LOCld}.b the I$^2$C configuration and the control circuits
for the total current passing through the M7 are illustrated.
The driver components M1-M8 are NMOS and PMOS devices, which are sensitive to
TID induced defects~\cite{MPKing}.

\begin{figure}[p!] 
  \vspace{-2mm}    
  \centering
    \includegraphics[height=120pt]{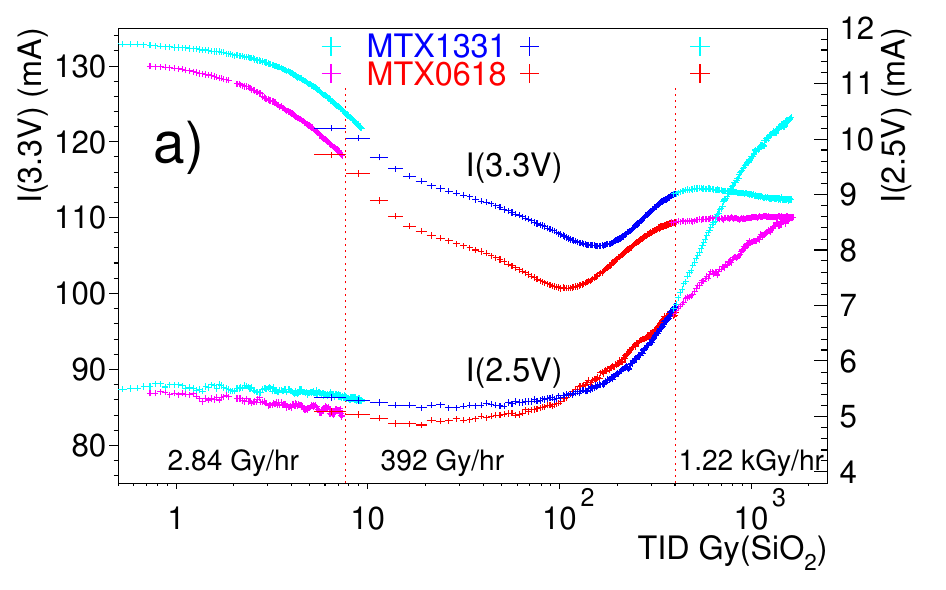}

    \vspace{-2mm}
    \includegraphics[height=120pt]{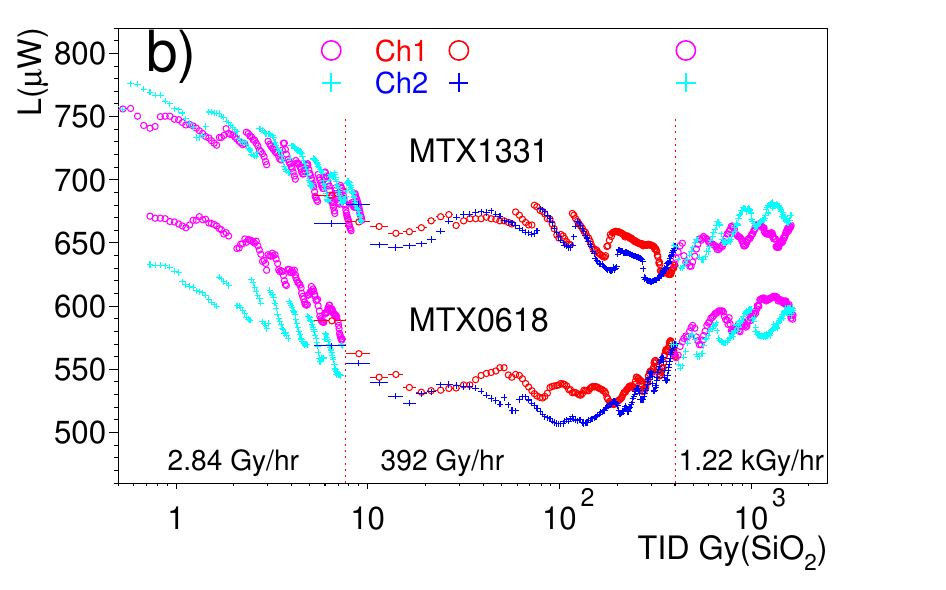}
    \vspace{-4mm}    
  \caption{ Two MTx samples were irradiated in X-ray at increasing dose rates, and  
  accumulated to {\color{black} 1.62 kGy(SiO$_2$).}
  The currents of the two bias voltages (3.3~V and 2.5~V) are plotted in a). 
  The 3.3~V currents dropped instantly at initial TID, to the lowest level at around 
  {\color{black} 100~Gy(SiO$_2$),}
  and recovered afterwards. The 2.5~V currents increased significantly 
  after {\color{black} 100 Gy(SiO$_2$).}
  The optical powers plotted in b) correspond to the 3.3~V currents, 
  and recovered to about 90~\% of the original. 
  The {\color{black} jagged curves} represent the characteristic changes in the ASIC for 
  the output currents to VCSELs.
  \label{fig:Xray-plots} }

  \centering
    \includegraphics[width=0.62\linewidth]{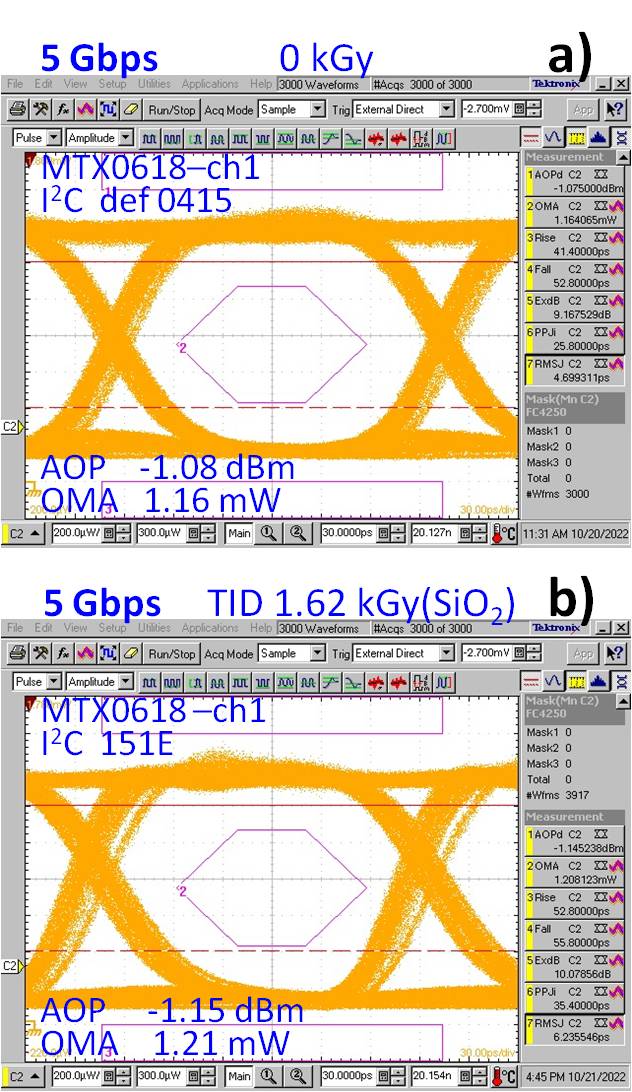}
  \caption{ Eye diagrams of an MTx optical output channel at 5 Gbps, 
    a) before and b) after the X-ray TID of
    {\color{black} 1.62 kGy(SiO$_2$).}
    At the default configuration (I$^2$C register value 0x0415),
    the 3.3 V current, AOP and OMA after TID dropped to 85~\% of the original. 
    In b), the I$^2$C setting (0x151E) has tuned values to 
    increase the AOP and OMA. However, the jitter noise 
    (increased by 25~\%) persists.
  \label{fig:Xray-eye} }
\end{figure} 

\section{ MTx irradiated in X-ray  }

The X-ray irradiation was conducted in an X-RAD (iR160) 
radiator\footnote{ \color{black} Precision X-Ray Inc., Madison, CT, USA.}
that has the photon energy distributed in $10 - 160$~keV.
{\color{black}
The radiation dose rate was calibrated with an Unfors Xi Transparent 
Detector\footnote{
\color{black} Unfors RaySafe AB, Billdal, Sweden.}.
The photon energy spectrum in air was measured with a 2-mm aluminum filter.
The absorbed dose in SiO$_2$ was converted according to the 
mass-energy absorption coefficients of air and SiO$_2$~\cite{Ravotti}.
The dose rate uncertainty is estimated to be 6 \%
due to the geometrical offset and 
the photon energy distribution.}

\begin{figure*}[t] 
  \vspace{-4mm}
  \centering
    \includegraphics[width=0.7\linewidth]{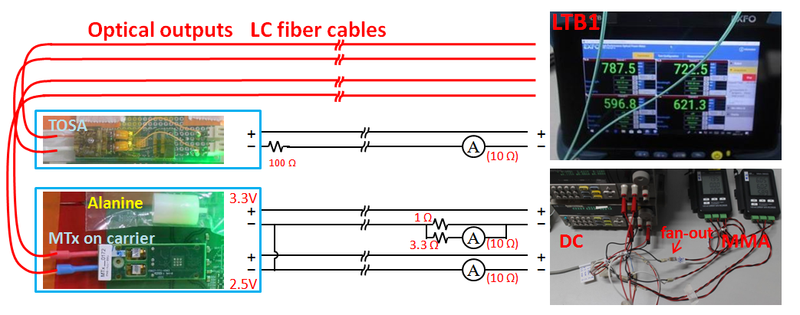}
  \vspace{-3mm}    
  \caption{ The Co-60 test setup is illustrated for the equipment 
  connected with 40 m cables and fibers to the VCSEL and MTx samples. 
  The test samples were attached with an Alanine dosimeter for TID calibration. 
  The VCSEL optical powers were recorded by a four-channel meter (LTB1, EXFO). 
  The currents of power supplies to TOSA and MTx 
  samples were recorded with current meters (MMA-386SD).
  The parallel resisters to the 3.3~V ground were applied to extract
  a small fraction of current 
  for the meter dynamic range limited to 20 mA.
  \label{fig:Co60-wiring} } 
\end{figure*} 

The MTx was prepared with the bias currents and optical powers recorded online.
Two MTx samples were tested in sequence with the
same dose rates and periods. The accumulated TID is {\color{black} 1.62~kGy(SiO$_2$).} 
Plotted in Fig.~\ref{fig:Xray-plots} are the observables in consecutive doses.
The irradiation started at the lowest tunable rate of 
2.84~Gy/hr and lasted 
for 160 minutes to  7.59~Gy. 
Afterwards the rate was raised to   392~Gy/hr  and 
then to 1.22~kGy/hr  for an hour each. 

The 3.3 V bias currents degraded rapidly by 25~\% at around 100 Gy, 
and recovered partially with additional TID to 85~\% of the original.
The optical powers have similar characteristics and recover to 90~\% of 
the initial level. 
Although the 2.5 V currents increased significantly
by about 50~\%, 
the I$^2$C communications of both modules were functional. 

The eye diagrams after irradiation have slightly increased jittering parameters.
Plotted in Fig.~\ref{fig:Xray-eye} are the 5~Gbps eye diagrams of a module 
before and after the X-ray TID of 1.62~kGy.
With the default configuration, 
the Optical Modulation Power (OMA) and Average Optical Power (AOP) 
are lower by about 15~\% after irradiation. 
By increasing the modulation voltage 
(Vmod in Fig.~\ref{fig:LOCld}.b from 6.4 to 6.8 V),
laser current (from 6 to 7~mA), and peaking strength (Vpk),
the eye diagram (Fig.~\ref{fig:Xray-eye}.b) is tuned with
the AOP and OMA closer to the original before irradiation.

\section{ MTx irradiated in Co-60  }

The Co-60 irradiation of MTx was conducted 
at the Institute of Nuclear Energy Research (INER), where
the gamma-ray facility has the Co-60 pellets of 10 mm diameter 
arranged in an array of $45 \times 300$ cm$^2$, which provides
a wider and uniform gamma ray distribution.
{\color{black}
The Co-60 test setup is illustrated in Fig.~\ref{fig:Co60-wiring}. 
The MTx samples in irradiation had an Alanine pellet closely attached. 
The absorbed dose in Alanine was measured by an electron paramagnetic resonance 
(EPR) analyzer\footnote{\color{black}
The AWM230 Alanine EPR Dosimeters were measured by a  
Bruker EMS 104 EPR analyzer.}.
The uncertainty is dominated by the geometric uniformity 
estimated to be 3~\%.
 
The Co-60 emits gamma-rays of 1.1 and 1.3~MeV.
With the photon energy larger than 120~keV, the mass-energy absorption coefficients
are approximately equal for Alanine and SiO$_2$.
The dose-conversion factor applied is 1~Gy(SiO$_2$) = 0.93 Gy(Alanine)~\cite{Ravotti}.
}

The MTx modules in tests were kept at $22\pm4$~$^\circ C$.
By the distance of test objects to the source, 
with additional Pb shielding, the dose rates were configured
from {\color{black} 0.14 to 45~Gy(SiO$_2$)/hr.}
The location of to power supplies and readout electronics 
is 40 m away in the external safe area.  
The data acquisition was recording every minute with a 4-channel 
optical power meter (LTB1) and current meters (MMA-386SD).
The irradiation was conducted daily during working hours. 
The data taking continued for source-off hours to detect 
annealing behaviors.

\begin{figure}[b!] 
   \vspace{-2mm}     
  \centering          
    \includegraphics[height=120pt]{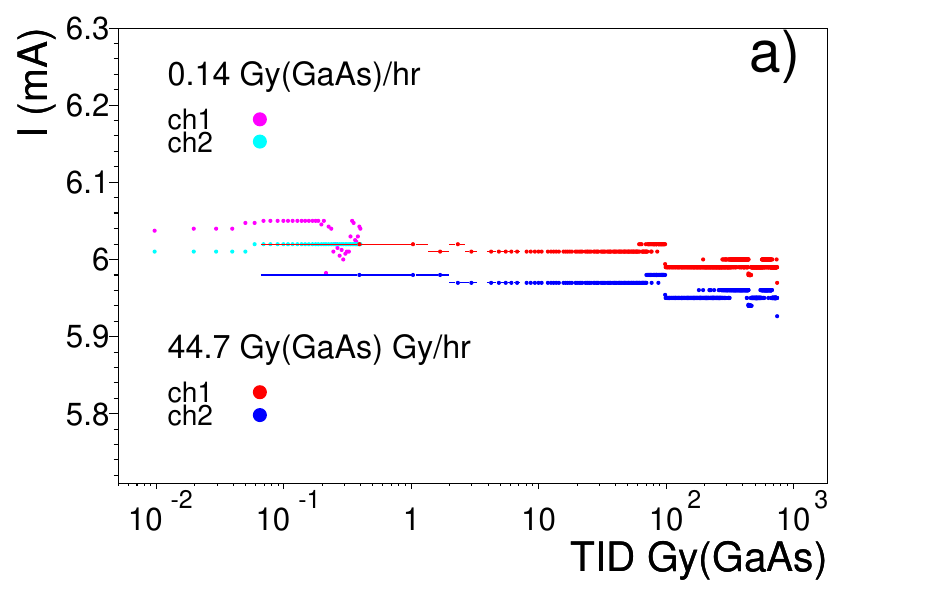} 
    \includegraphics[height=120pt]{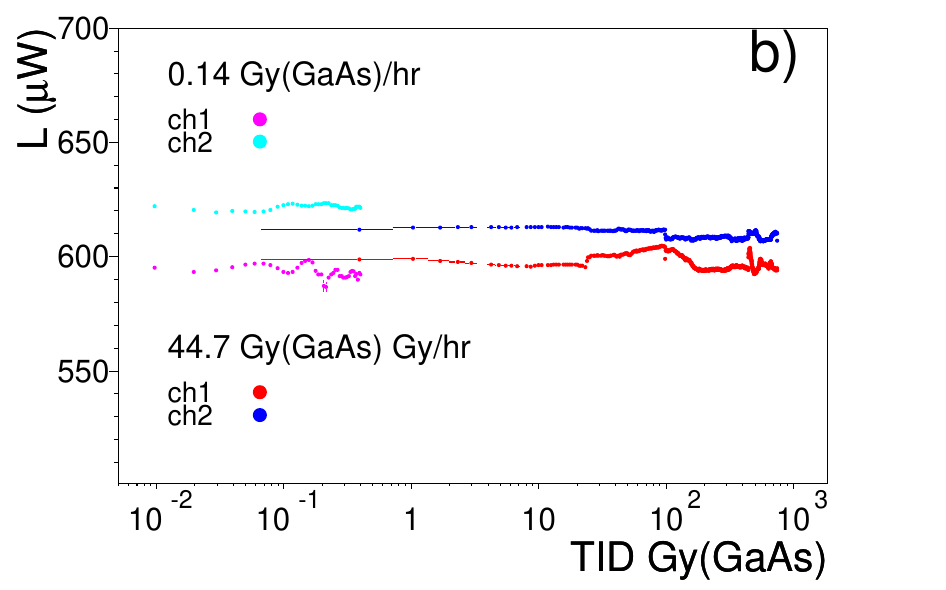}
    \vspace{-3mm}    
  \caption{ Two of the TOSA packaged VCSELs (TTF-1F59, Truelight) were irradiated with Co-60
  at constant currents of 6~mA,  initially at {\color{black}
  0.14~Gy(GaAs)/hr to 0.38~Gy(GaAs)},
  and then at {\color{black} 44.7~Gy(GaAs)/hr to 1.59~kGy(GaAs).}
  The currents and optical powers are 
  plotted. The high rate is shown for the latest period of {\color{black}
  0.74~kGy(GaAs).}
  The steps were caused by accesses before source-on of working days.
  \label{fig:TOSA} }
\end{figure} 

{\color{black}
The VCSELs fabricated in GaAs substrate were tested independently 
for the TID effects. The GaAs mass-energy absorption coefficient 
in the Co-60 photon spectrum is about equal to the SiO$_2$ \cite{DoseConversion}.
The same dose conversion factor is applied to the Alanine measurement.
}

The TOSA packaged VCSEL samples (TTF-1F59) 
were connected to power supplies at constant currents of 6 mA.
The irradiation started at  0.14~Gy/hr 
to accumulate  5.77~Gy, 
and then at  44.7~Gy/hr to a total {\color{black} 1.59~kGy(GaAs).}
In Fig.~\ref{fig:TOSA}  the TOSA currents
and optical powers are plotted for periods of the two dose rates. 
The stable distributions suggest that the VCSELs are not affected by TID.

The degradation of MTx in TID is attributed by the characteristic changes
in the LOCld.
The irradiation initiated at 0.14~Gy/hr,
with three MTx samples exposed to various total doses of 
0.37 to {\color{black} 3.81~Gy(SiO$_2$).}
The reductions on currents and optical powers were instant, 
and deviated largely from 10~\% to 15~\%.

\begin{figure}[b!]     
  \centering
    \includegraphics[height=125pt]{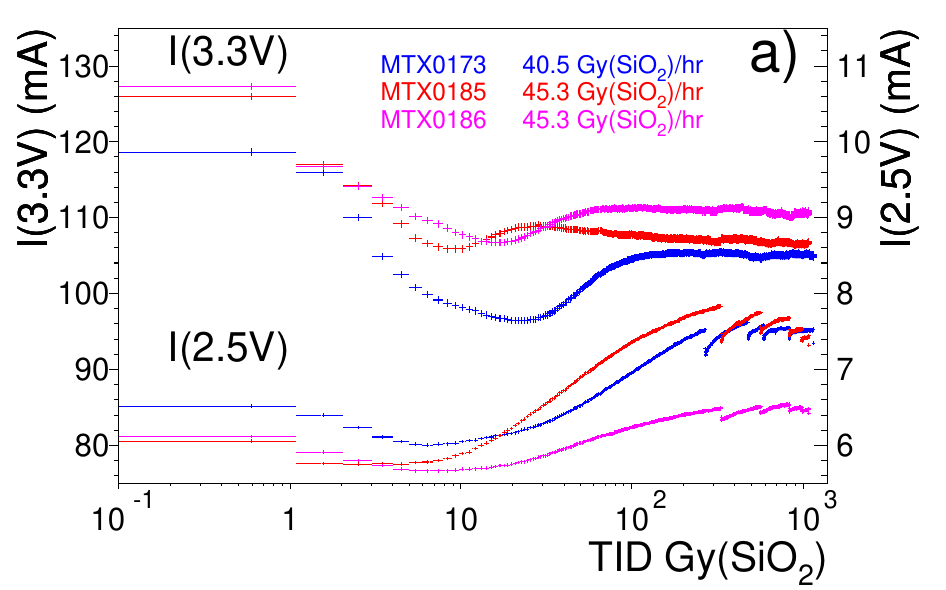}
    \includegraphics[height=125pt]{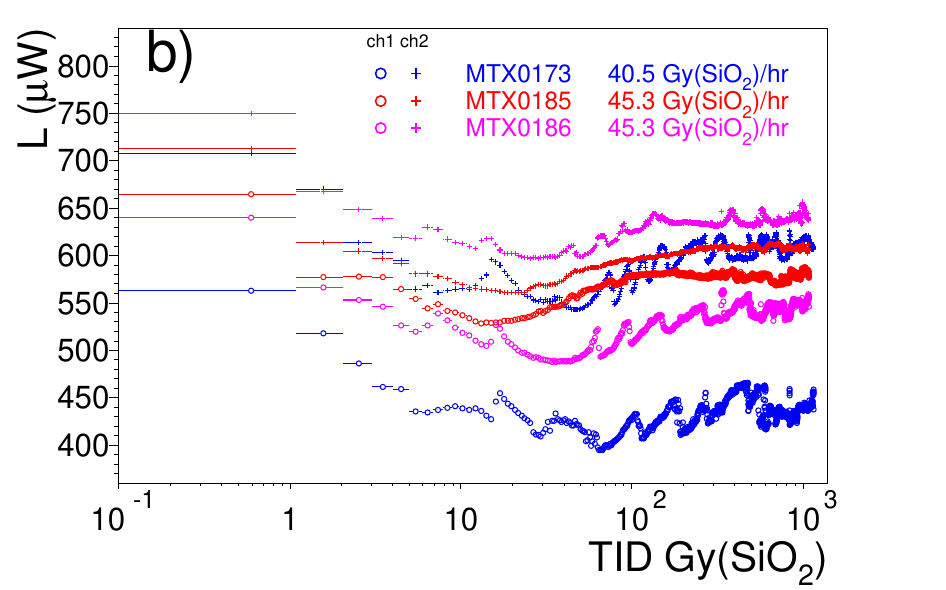}
    \vspace{-4mm}    
  \caption{ The a) bias currents and b) optical powers are plotted for three MTx’s 
  irradiated in Co-60 gamma-rays at  
  {\color{black} 44.6~Gy(SiO$_2$)/hr to  1.15~kGy(SiO$_2$).}
  The breaking steps in 2.5~V currents correspond
  to the annealing overnight to lower currents at the beginnings of 
  source-on in each working day.
  The 3.3~V currents and optical powers degraded to the lowest at around 10 to 50~Gy,
  and recovered afterwards to 90~\% of the original.
  \label{fig:Co60TID} }
  \vspace{-2mm}
\end{figure} 

\begin{figure}[bh!] 
  \vspace{-5mm}
  \centering
    \includegraphics[width=1.05\linewidth]{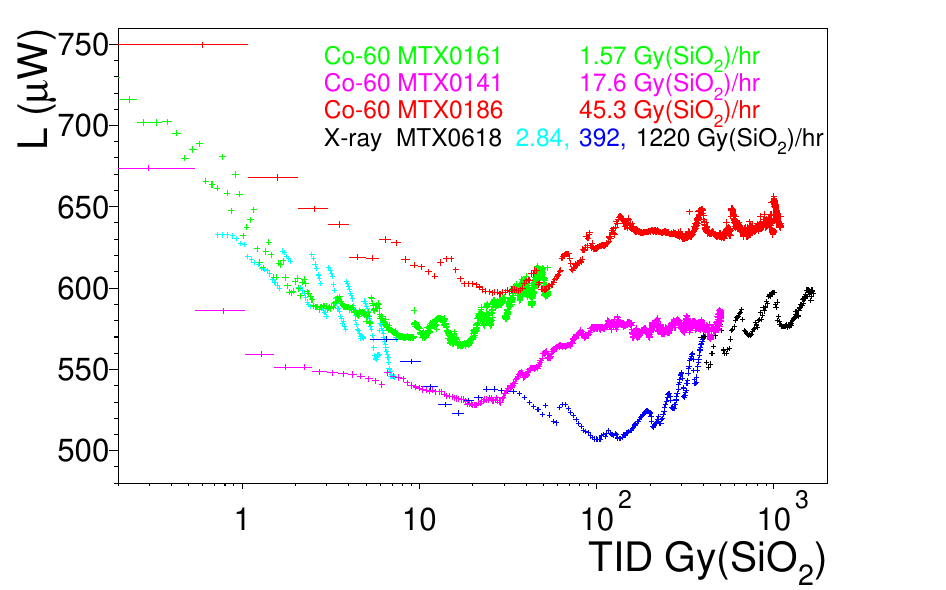}
    
    \vspace{-3mm}    
  \caption{  \color{black} The laser light powers are shown for the irradiation 
  with Co-60 at dose rates of 1.57 to 45.3 Gy(SiO$_2$)/hr, and with X-ray at increasing rates.
  The degradation and recovery occurred sooner with lower Co-60 dose rate.
  The X-ray irradiation
  was conducted without annealing, the recovery occurred around 100 Gy(SiO$_2$).  
  \label{fig:rate-dep} }

   \vspace{2mm}
  \centering
    \includegraphics[height=125pt]{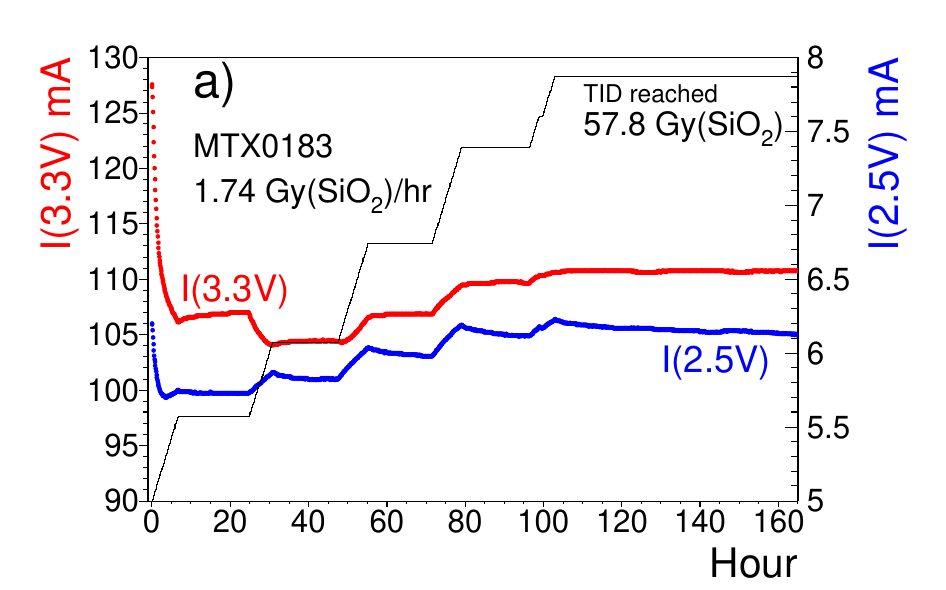}

    \vspace{-4mm}
    \includegraphics[height=125pt]{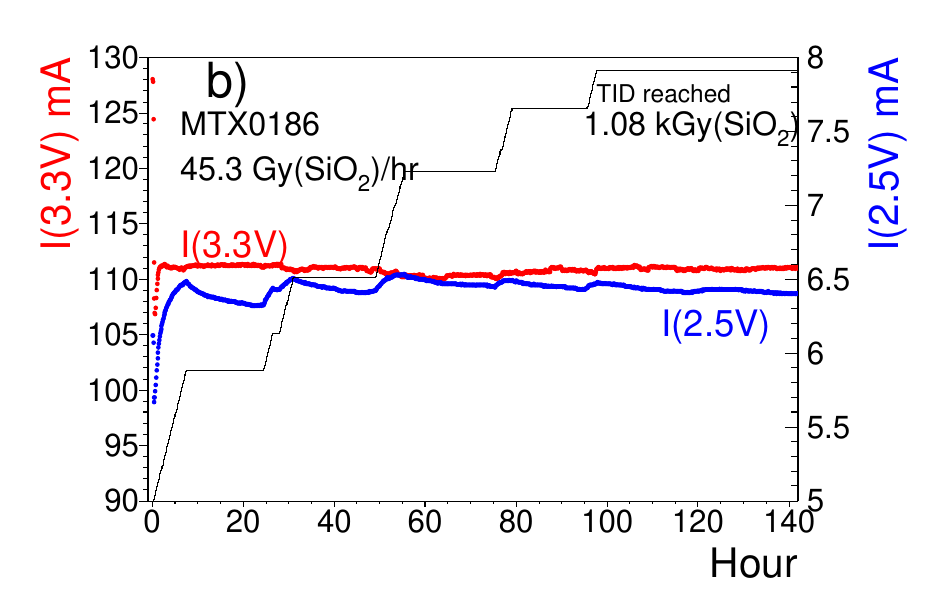}
    \vspace{-3mm}    
  \caption{\color{black} The currents in time of Co-60 irradiated samples at the dose rates
  of 1.74 and 45.4~Gy(SiO$_2$)  are illustrated in a) and b), respectively. The accumulated 
  TIDs in source-on and -off periods are indicated by the black lines, in five working days.
  The degradation of currents occurred quickly at the initial 10~Gy(SiO$_2$).
  The annealing is more effective at lower dose rate, and at the initial TID. 
  The recovery in currents occurred before 100~Gy(SiO$_2$).
  \label{fig:Co60-annealing} }
  \vspace{-4mm}
\end{figure} 

To accelerate the total dose accumulation, and for comparison of the dose rate 
dependence, the MTx samples were irradiated at 1.7, 18 and 45~Gy/hr. 
Plotted in Fig.~\ref{fig:Co60TID} are the measurements of samples 
tested at 45~Gy/hr, to a total of {\color{black} 1.15~kGy(SiO$_2$).}
The responses observed are similar to those tested at lower rates.
The currents and optical powers degraded with the initial doses of
10 to 50~Gy, and recovered partially with additional TID.

{\color{black}
The two VCSEL diodes of a MTx, driven at $\sim $6~mA, are
biased by the 3.3 V that has a total current of 120~mA.
The jagged curves of the optical outputs reflect the effects 
due to buildup of charges and defects 
in the driver circuits and current regulation, 
biased at 3.3 V and 2.5 V, respectively.
The rate dependence is illustrated in Fig.~\ref{fig:rate-dep},
with the optical powers measured at four dose rates.
At higher dose rates, the recovery is observed shifting
from 10 Gy at 1.57 Gy/hr,
to 100 Gy with the X-ray at 392~Gy/hr.
  
The annealing effect is visible on the 2.5~V currents.
The LOCld bias currents measured in time are plotted 
in Fig.~\ref{fig:Co60-annealing}, for the dose rates of a) 1.74 
and b) 45.3 Gy/hr, respectively.
The doses accumulated are also plotted, to indicate
the degradation and annealing during source-on/off hours, respectively.
At low dose rate the recovery occurred around 20 Gy.
The annealing is more effective in the initial doses.
}

\begin{figure}[b!] 
  \vspace{-4mm}
  \centering
    \includegraphics[width=0.66\linewidth]{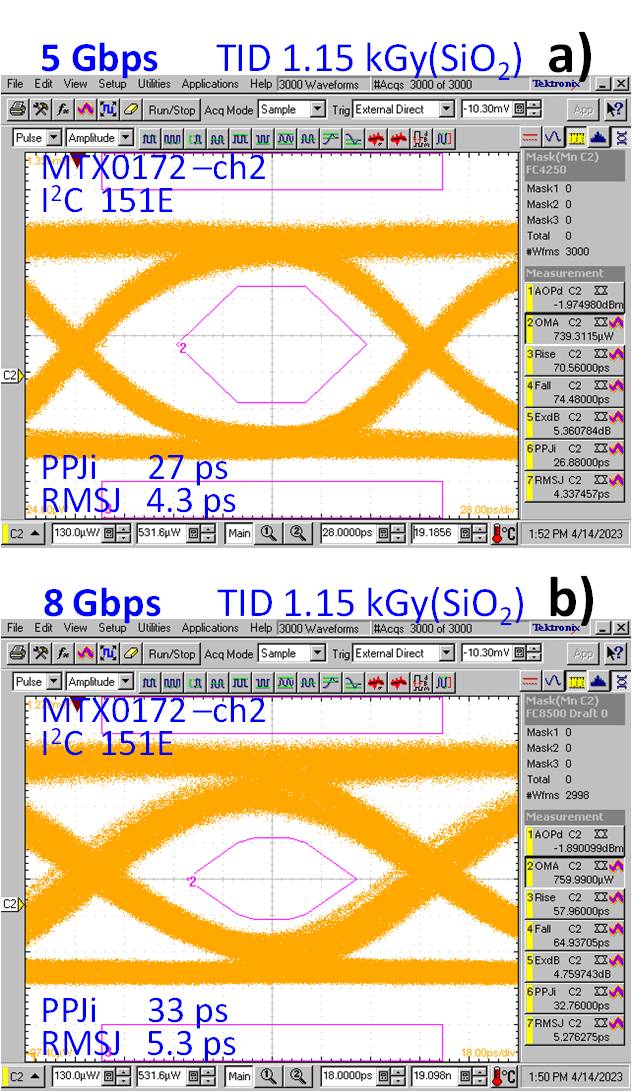}
  \caption{ The eye-diagrams measured at a) 5 Gbps, and b) 8 Gbps are 
            plotted for an MTx channel irradiated with Co-60 to 
            {\color{black}  1.15~kGy(SiO$_2$).} 
            The LOCld configuration (I$^2$C register 0x151E)
            is optimized for higher optical power. The Peak-Peak (PPJi) {\color{black}
            and RMS jitter parameters are larger by about 20~\% 
            with respect to the initial values before irradiation.} 
  \label{fig:Co60-eye} }
\end{figure}

In total fifteen MTx modules were tested in Co-60 irradiation.
{\color{black}
Each was conducted at a constant dose rate (between 0.14 to 45.3 Gy/hr), 
with the accumulated TIDs of 0.37 to 1.15~kGy. }
The eye diagrams of them were examined at 5 and 8~Gbps speeds.
Plotted in Fig~\ref{fig:Co60-eye} are the eye diagrams of a  
MTx channel exposed to  1.15~kGy. 
The LOCld is configured with the parameters used for 
the X-ray irradiated samples (Fig.~\ref{fig:Xray-eye}.b), 
to enhance the AOP and OMA magnitudes. {\color{black}
The jitter parameter values are higher than the original by about 20~\%.}
All the irradiated samples can pass 8~Gbps bit-error rate tests.
 
\section{Summary }
\label{sec:summary}

The radiation tolerance of the MTx transmitter in TID has been studied
with X-rays to {\color{black} 1.62~kGy(SiO$_2$) }
and with Co-60 gamma-rays to {\color{black} 1.15~kGy(SiO$_2$). } 
The degradation of LOCld output currents and optical powers occurs at
initial TID to the lowest level at around {\color{black} 10 to 100~Gy(SiO$_2$),}
and recovers with increasing TID, to 90~\% of the original.
The results with X-rays and Co-60 gamma-rays are compatible.
The 5 and 8~Gbps eye diagrams of Co-60 irradiated samples
show similar increases on jitter parameters by about 20~\%.  
The data transmission is qualified for the ATLAS 
applications at 5~Gbps speed.

\vspace{-2mm}
\section*{Acknowledgement}

The authors would like to thank the assistance of the Institute of Nuclear Energy Research.
This work has been partially supported by the Institute of Physics, Academia Sinica,
and the Ministry of Science and Technology, grant MOST 1082112-M-001-047. 

\vspace{-2mm}

{}


\begin{thebibliography}{}
\vspace{-2mm}

\bibitem{MTx}  
C. Liu, et al., The miniature optical transmitter and transceiver
for the High-Luminosity LHC (HL-LHC) experiments, JINST 8 (2013) C12027.

\bibitem{MTx_MTRx}  
X. Zhao, et al., Mid-board miniature dual channel optical transmitter MTx and
transceiver MTRx, JINST 11 (2016) C03054.

\bibitem{MTx_trigger} 
L. Zhang, et al., Optical transceivers for event triggers in the ATLAS phase-I upgrade,
Nucl. Instrum. Methods A985 (2021) 164651.

\bibitem{LAr} 
ATLAS Collab, ATLAS Liquid Argon Calorimeter Phase-I Upgrade Technical
Design Report, ATLAS-TDR-022, 2013.

\bibitem{NSW} 
ATLAS Collab, ATLAS New Small Wheel Technical Design Report, ATLAS-TDR-020, 2013.

\bibitem{GBTIA} 
M. Menouni, et al., The GBTIA, a 5 Gbit/s radiation-hard optical receiver for
the SLHC upgrades, in: Proceedings of TWEPP-09, CERN-2009-006, 2009.


\bibitem{LOCld} 
F. Liang, et al., The design of 8-Gbps VCSEL drivers for ATLAS
liquid Argon calorimeter upgrade, JINST 8 (2013) C01031.

\bibitem{MTx_prottype}  
B. Deng, et al., Component prototypes towards a low-latency, small-form-factor
optical link for the ATLAS Liquid Argon Calorimeter Phase-I trigger upgrade,
IEEE Trans. Nucl. Sci. 62 (2015) 250.

\bibitem{Oldham}{\color{black}
T. R. Oldham and F. B. McLean, Total ionizing dose effects in MOS oxides and devices,
IEEE Trans. Nucl. Sci. 50 (2003) 483.}

\bibitem{Opto-radhard} 
S. Hou, et al., Radiation hardness of optoelectronic components for the optical
readout of the ATLAS inner detector, Nucl. Instrum. Methods A636 (2011) S137.

\bibitem{MTx_Aging} 
F.X. Chang, et al., Aging and environmental tolerance of an optical transmitter
for the ATLAS Phase-I upgrade at the LHC, Nucl. Instrum. Methods A831 (2016) 349.

\bibitem{peaking}
F.-T. Liang, et al., Active inductor shunt peaking in high-speed VCSEL driver design,
Chin. Phys. C 37 (2013) 116101.

\bibitem{LOCld_ASIC} 
X. Li, et al., 8-Gbps-per-channel radiation-tolerant VCSEL drivers for the LHC
detector upgrade, JINST 10 (2015) C02017.


\bibitem{MPKing}
M.P. King, et al., Response of a 0.25 $\mu$m thin-film silicon-on-sapphire
CMOS technology to total ionizing dose,
JINST, 5 (2010) C11021.  

\bibitem{Ravotti}{\color{black}
F. Ravotti, Dosimetry Techniques and Radiation Test Facilities for Total Ionizing Dose Testing,
IEEE Trans. Nucl. Sci. 65 (2018) 1440.}

\bibitem{DoseConversion}{\color{black}
L.D. Edmonds, Dose Conversions between Different Materials,
DOI:10.13140/RG.2.1.3145.7449, (2015).}








\end{thebibliography}
\end{document}